\documentclass[a4paper,fleqn]{cas-dc}
\usepackage{amsmath}
\usepackage{mathtools}

\usepackage[utf8]{inputenc}
\usepackage[numbers]{natbib}
\usepackage[linesnumbered,ruled]{algorithm2e}

\DeclareMathOperator*{\argmax}{arg\,max}
\def\tsc#1{\csdef{#1}{\textsc{\lowercase{#1}}\xspace}}
\tsc{WGM}
\tsc{QE}
\tsc{EP}
\tsc{PMS}
\tsc{BEC}
\tsc{DE}

\begin{document}
\let\WriteBookmarks\relax
\def\floatpagepagefraction{1}
\def\textpagefraction{.001}
\shorttitle{Tales of a City: Sentiment Analysis of Urban Green Space in Dublin}
\shortauthors{M. Ghahramani et~al.}

\title [mode = title]{Tales of a City: Sentiment Analysis of Urban Green Space in Dublin}                      

\author[1]{Mohammadhossein Ghahramani}
\cormark[1]
\ead{sepehr.ghahramani@ucd.ie}

\author[1,2]{Nadina J. Galle}
\ead{nadina.galle@ucd.ie}

\author[2]{Carlo Ratti}
\ead{ratti@mit.edu}

\author[1]{Francesco Pilla}
\ead{francesco.pilla@ucd.ie}


\address[1]{Spatial Dynamics Lab, University College Dublin, Ireland}
\address[2]{Senseable City Laboratory, Massachusetts Institute of Technology, USA}

\cortext[cor1]{Corresponding author}

\begin{abstract}
Social media services such as TripAdvisor and Foursquare can provide opportunities for users to exchange their opinions about urban green space (UGS). Visitors can exchange their experiences with parks, woods, and wetlands in social communities via social networks. In this work, we implement a unified topic modeling approach to reveal UGS characteristics. Leveraging Artificial Intelligence techniques for opinion mining using the mentioned platforms (e.g., TripAdvisor and Foursquare) reviews is a novel application to UGS quality assessments. We show how specific characteristics of different green spaces can be explored by using a tailor-optimized sentiment analysis model. Such an application can support local authorities and stakeholders in understanding—and justification for—future urban green space investments.
\end{abstract}

\begin{keywords}
Smart Cities \sep Urban Green Space \sep Natural Language Processing \sep Urban Planning \sep Artificial Intelligence
\end{keywords}

\maketitle

\section{Introduction}
The lack of quality assessments underestimates the value of urban green space (UGS) due to a lack of information about what quality green space entails and how existing spaces within a city score on important social quality parameters \cite{CIMBUROVA2020126801}. Measuring the quality of urban green space is a tedious process. Observational approaches are often criticized since they require extensive repeat measurements at the same location, incurring considerable time and cost expenditures. Moreover, collected data quickly become outdated, making progress out of reach \cite{Hu2020Spatial}. Since green space can be a powerful ecological instrument and a possible way to improve public health, city planners, urban designers, policymakers, and estate managers must maximize opportunities to apply greening strategies. Since increasing urban greening is unattainable, the focus should be on accelerating the quality of urban nature. But how can a city ensure that it provides safe, inclusive, and high-quality green space for all? Several attempts have been made to streamline and standardize quality assessments of urban green space \cite{VANO2021126957,LI2020126903}. Still, most proposed methods fall short, as they require expensive, tedious, and time-consuming field surveys and in-situ observations. 

AI has made its way into different areas, e.g., engineering, urban planning, management, and manufacturing to solve various problems \cite{Ghahramani-Spatial,Ghahramani-Community,SemiConductorGH}. Sentiment analysis \cite{GHAHRAMANI2021100058} can offer a way to make evidence-based improvements for urban green space. Utilizing social media platforms such as TripAdvisor to mine people's green space opinions can open up a treasure trove of global, comparative data. To date, no other crowdsourced data has been widespread enough to allow for such comparison between green spaces, both within a city and beyond. Thus far, sentiment analysis using online platforms as a data source has only been applied in the hospitality and tourism sectors. Here, shallow NLP techniques are applied to extract sentiment automatically. This paper uses two social media platforms to address this need and translate their reviews about urban green space into relevant, easily understandable insights for cities. The goal is to ''mine'' the reviews for opinions and extract the sentiment behind them by leveraging Artificial Intelligence (AI) \cite{Ghahramani2020AI1,Ghahramani2020AI2,GhahramaniCovid} and Natural language processing (NLP) \cite{Ghahramani2020Urban,SERNA20171}. Such quantification can provide city planners, policymakers, and landscape architects an invaluable tool for understanding and justifying future investment in urban green space \cite{ACHILLAS2011414,PANEK201765,Carbon2021}.

In this work, UGS in Dublin, Ireland, is analyzed by capturing and exploring public opinion. A tailor-optimized topic modelling method is implemented to classify UGS reviews and discover latent themes automatically. We discuss how social media platforms (e.g., Foursquare and TripAdvisor) and AI-based techniques can be utilized to assess the quality of urban green space cost-effectively. This paper presents an experiment's results to use NLP to extract citizen opinion on UGS quality, a novel application of automatic text mining and sentiment analysis. We present how text mining and sentiment analysis can help authorities understand how people value and use urban green spaces. Such intelligent approaches can help policymakers understand the needs and problems associated with UGS and allow them to make better decisions.

\section{Sentiment Analysis}\label{Method}

\subsection{Urban Green Space in Dublin}
\begin{figure}
  \includegraphics[width=\linewidth]{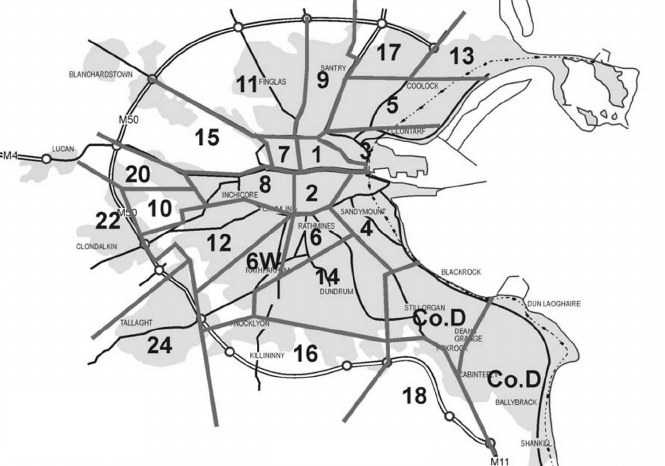}
  \caption{Dublin's postal districts}
  \label{dublinDistricts}
\end{figure}

UGS such as parks, woods, and wetlands represent a fundamental component of Dublin's urban ecosystem. It has become an extension of, or in many cases, replacing the traditional backyard, meaning more people are sharing less green space. Dublin City is divided into different postal districts illustrated in Fig. \ref{dublinDistricts}. Some of them (the odd numbers) are located on the North-side of the River Liffey, and the rest (the even numbers) are situated on the South-side. There are various green areas, i.e., parks, scattered throughout all these districts.

Social network platforms can enable us to utilize the abundance of open-source reviews. However, Dublin follows a similar pattern as other cities, where the most popular parks have significant reviews, and the lesser common parks have significantly fewer reviews. Therefore, we consider the most popular parks as a proxy for UGS. As discussed, this paper aims to utilize social network platforms and AI techniques to quantify UGS quality across Dublin based on a hybrid model. First, a location intelligence-driven technology (i.e., Foursquare API) is employed to find parks across Dublin, and the popularity of each park is assessed. Such a popularity measure is aggregated given all postal districts illustrated in Fig. \ref{dublinDistricts}. Then, associated reviews of all popular parks are extracted from the TripAdvisor platform. Then, different opinions clusters are extracted to identify related characteristics of parks. Relevant details are explained next.

\begin{figure}
  \includegraphics[width=\linewidth]{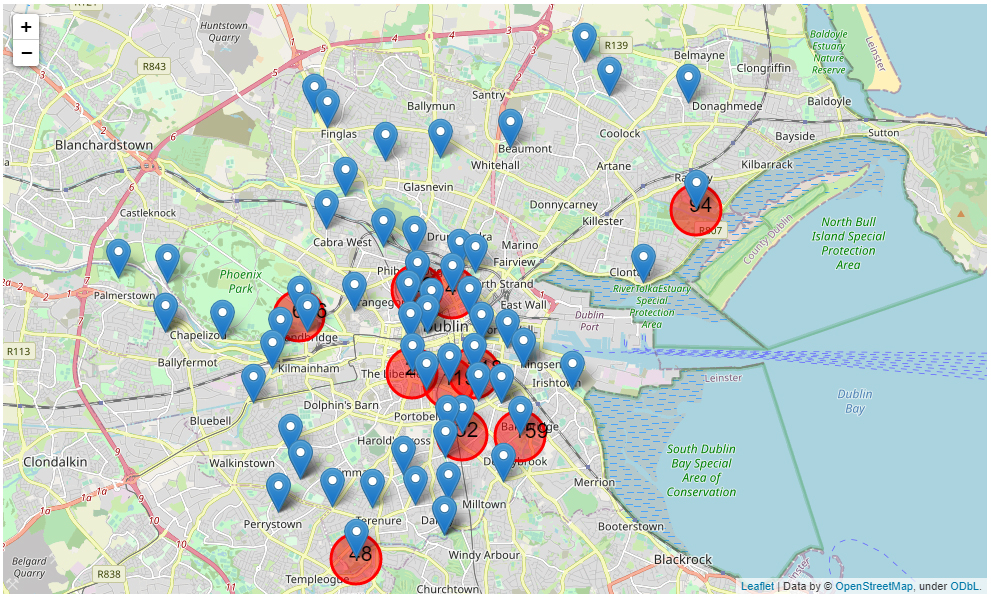}
  \caption{Distribution of identified parks in Dublin city given Foursquare API.}
  \label{parksDublin}
\end{figure}

\subsection{Extracting Dublin's parks data}
\begin{figure}
  \includegraphics[width=\linewidth]{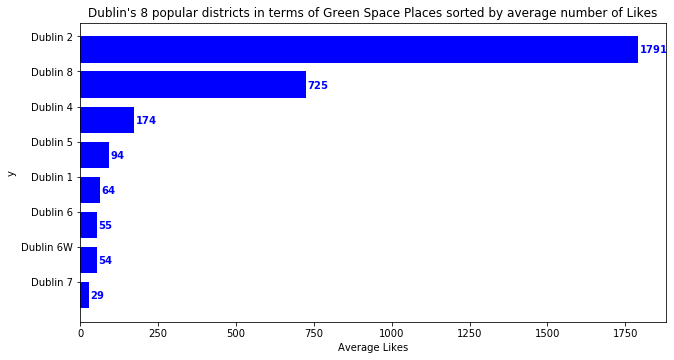}
  \caption{Dublin's 8 popular districts as to Green Space Places.}
  \label{parkLikes}
\end{figure}

Foursquare API provides access to various capabilities, such as location search and venue details. The API has two different types of endpoints, i.e., regular and premium, each providing access to various information. It uses two forms of authentication, i.e., Userless Auth and User Auth. The former is used for server-side applications, while the latter uses OAuth 2.0 to provide authorized access to the API. In this work, we have used the Userless Auth authentication form. This authentication type requires the consumer key's Client ID and Secret instead of an access token in the request URL. Various information, such as location data, i.e., street address, postal code, longitude, latitude, and distance, can be fetched. We found around $60$ parks across Dublin (Fig. \ref{parksDublin}) and aggregated them at the postal district level. Some examples of such distribution are presented in Table \ref{number-parks}. The API also provides access to information about each venue's level of popularity, i.e., the counts of users who have liked a venue. We used such capability to assess the popularity of all parks detected. Each observation (i.e., park) includes a unique string identifier, the postal district it belongs to, and the number of likes. The pseudo-code for the procedure is presented in Algorithm \ref{like-model}. Given the procedures described in the algorithm, the number of likes for each park is revealed (Table \ref{parksLikes}, 8 popular parks). Then, such measurements were aggregated at the postal district level. Most popular parks (together with their number of likes) are depicted in Fig. \ref{parksDublin}. Moreover, most popular districts evaluated based on the parks popularity measure using the Foursquare platform are also presented in Fig. \ref{parkLikes}.

\begin{table}[ht]
\caption{Some of Dublin's Districts, their coordinates and the number of parks in each area } 
\centering 
\scriptsize
\begin{tabular}{c c c c c c c c c c c c c c} 
\hline\hline 

  Districts & Area & Latitude & Longitude & No. Parks 
\\ [0.5ex]
\hline 
 
Dublin 1 & Northside &53.352488	&-6.256646	&7	    \\[0.5ex]

\hline 

Dublin 2	&Southside	&53.338940	&-6.252713	&8     \\[0.5ex]
\hline 
 
Dublin 3	&Northside	&53.361223	&-6.185467	&3	
    \\[0.5ex]

\hline 
 
Dublin 4&	Southside	&53.327507	&-6.227486	&6
    \\[0.5ex]
\hline 
 
Dublin 5	&Northside	&53.383454&-6.181923	&3	
    \\[0.5ex]
\hline 
 
Dublin 6	&Southside	&53.317698	&-6.259525	&6
    \\[0.5ex]

\hline 
 
Dublin 6W	&Southside	&53.309282	&-6.299435	&4	
    \\[0.5ex]
 \hline 

\end{tabular}
\label{number-parks}
\end{table}

\begin{algorithm}
\scriptsize
\SetAlgoLined

    \SetKwInOut{Input}{Input}
    \SetKwInOut{Output}{Output}
    \Input{$Dublin_d \gets$ Postal districts' coordinates;\\ $C_{id}\gets \text{Client ID}$; $C_{s}\gets \text{API Secret}$;\\
    $Limit \gets limit_0$;\\
     $URL \gets \text{https://api.foursquare.com/v2/venues/search?+queries}$\\
    }
    
    \Output{Parks' information}
    \# \enspace \textit{Retrive Function} \\
        {Parks = request.get(URL) }\;

        \For{$i \gets Parks$}{
        \If{$i.startswith(location)$}
    	{ $return \text{ }{ i},i_{id}, i_{name};$}
		}
    $Data \gets []$;\\                    
     $k \gets 0$;\\              
       \For{$i,j \gets Dublin_d.iterrows()$}{
    	{ $X=Retrive(lat=j[lat],lon=j[lon],search\_query=Park, radius=1200);$}
    	 { $X[districts]=Dublin_d.at[i,districts];$}\\
    	 \uIf{k !=0}{
    $Data=Data.append(X);$ \\
  }
  \Else{
    $Data=X$; \\
    $Data[districts]=Dublin_d.at[i,districts]$; \\
  }
 $k \gets 1$;
		}   
	
      \For{$i,j \gets Data.iterrows()$}{
      $venue_{id} = j[id]$\\
      $result \text{ = } requests.get(URL)$\\
      $Data.at[i , Likes] = Response(result_i[likes])$
      
      }	
		
    {Output the result.}
    \caption{API querying for retrieving data}
    \label{like-model}
    \end{algorithm}

\begin{table}[ht]
\caption{8 detected most popular parks using the Foursquare platform} 
\centering 
\scriptsize
\begin{tabular}{l l c c c c c c c c c c c c} 
\hline\hline 

  \textbf{Name} &\textbf{ Area}  & \textbf{Likes}
\\ [0.5ex]
\hline 
 
St Stephen's Green & Dublin 2 	& 1191    \\[0.5ex]

\hline 

Phoenix Park &	Dublin 8		&696     \\[0.5ex]
\hline 
 
Merrion Square Park	& Dublin 2		&318	
    \\[0.5ex]

\hline 
 
St Patrick's Park	& Dublin 2		&267
    \\[0.5ex]
\hline 
 
Herbert Park	& Dublin 4		&159	
    \\[0.5ex]
\hline 
 
St Anne's Park \& Rose Gardens	& Dublin 5		&94
    \\[0.5ex]

\hline 
 
Mountjoy Square Park	& Dublin 1		&48	
    \\[0.5ex]
 \hline 

Bushy Park	& Dublin 6W		&48	
    \\[0.5ex]
 \hline 

\end{tabular}
\label{parksLikes}
\end{table}

\subsection{Extracting Reviews}
TripAdvisor reviews for 8 popular parks detected earlier were scarped using Selenium and Python. The pseudocode for retrieving data has been presented in our previous work \cite{GHAHRAMANI2021100058}. The reviews were subsequently processed to focus on English texts. The reviews were collected from the period May 2006 to December 2020. The following review fields were extracted: review-title (the written title of review); review-body (written review about the destination); rate-value (1 is the lowest evaluation, 5 is the best); review-location (where a reviewer is from); and review-date (date review was written). Some observations are presented in Table \ref{ObservationText}.

\begin{table*}[ht]

\caption{Examples of TripAdvisor reviews about  St. Stephen’s Green (Dublin, Ireland).} 
\centering 
\scriptsize
    \begin{tabular}{| c | p{13cm}  | l |}
    \hline
    Bubble\_rating & Review\_body & Date \\ \hline
    5 & Absolutely beautiful if you get the weather to enjoy it. Bring a picnic and enjoy the views. & 21-11-2018  \\ \hline
    4  & St. Stephen's Green park is perfect to step away from the hustle and bustle of Dublin. 
   & 12-06-2020 \\\hline
    1  &  Near the playground entrance there were about 200 teens drinking and causing trouble. & 20-08-2019 \\\hline
    2  & While we were walking across the park, a young man tried to take my husband's laptop. & 03-07-2019 \\
    \hline
    \end{tabular}
\label{ObservationText}
\end{table*}

\subsection{Topic Modeling}
When analyzing text data, we need to perform data cleansing operations to ensure the dataset's quality. Text normalization was used to convert text into more convenient, standard forms. Tokenisation was used to separate words from running text. Tokenised words, then, were converted into a numeric representation, a process known as vectorization. All punctuations and stop words were omitted. To that end, libraries from the Natural Language Toolkit were used. Each review text was converted into a list of words. A stemming algorithm was also used for reducing inflected words to their word stem by using suffix striping to produce stems. All URLs, email addresses, extra spaces, and emojis that existed in reviews were excluded. Bigrams and Trigrams were created, and lemmatization was performed.

The aim in this phase is to implement an optimized method to automatically extract what topics people are discussing about the most popular parks detected in the prior phase. The implemented approach is a variant of a Latent Dirichlet Allocation method tailored for this study to detect underlying topics \cite{GANGADHARAN20201337}. The contribution of each topic was also evaluated. This unsupervised learning approach enables us to discover hidden semantic structures in review comments. The method takes a collection of reviews and a few parameters as the input. It generates a probabilistic model so that the extent to which each review belongs to a given topic is revealed. The topics produced by the model are clusters of similar words in reviews. The model assumes that reviews can be represented as a probabilistic distribution over latent topics. Moreover, the association among all topics is analyzed. To infer the latent structure, we should represent reviews such that they can be manipulated mathematically. To do so, we represent each comment as a vector of features. A dictionary-based text categorization method was performed, a unique id for each word was created, and the output was converted to bag-of-words. In this way, the frequency counts of each word were determined. The vectorized corpus, then, transformed using a TF-IDF model. It transformed vectors from the bag-of-words representation to a vector space where the frequency counts were weighted given the scarcity of words.

The extracted review comments can be defined as a sequences of text, i.e., $X = \{X_1 , X_2, . . . , X_n\}$ where $X_i$ refers to the {$i^\text{th}$} review. Let $Z= \{z_1, z_2, ..., z_k \}$ be the set of $k$ topic models. We can define a log-likelihood objective function as 

\begin{equation}
\log P(X|Z) = \sum_{x\in X}^{} \log p(x| z_{\theta(x)})
\end{equation}
where $\theta(x) = \argmax\limits_{\theta} \log p(x| z_{\theta})$ is the topic identity of review $x$. The algorithm should maximize the defined objective function. 

A topic (cluster) was considered as a probability mass function over all the words. It has a probability of occurring from $0$ to $1$, and the sum of these probabilities is $1$. Moreover, each word was assigned an individual probability of occurring given a particular topic. Let $\Theta_{k,v}$ be the probability of a topic $k$ generating a word $v$. Let $\theta_{k}$ be the probability that review comments will generate a word from a specific cluster $k$. The values were generated by random variables, e.g., $\Theta_{k}$, with a Dirichlet distribution. Two different Dirichlet distribution functions were implemented, i.e., representing 1) how many words belong to clusters and 2) how associated clusters are with review comments. A Dirichlet distribution is a continuous multivariate density function with $2$ parameters, i.e., the number of topics ($k\geqslant2$) and concentration parameters ($ \alpha_k = \alpha_1, \alpha_2, ... \alpha_k$) \cite{OZYURT2021114231}. It outputs $k$ probabilities which sums up to 1, $\sum_{i=1}^{k} p(z_i)=1$ . The Dirichlet probability density function is

\begin{equation}
f(z_k) = \frac{\Gamma(\alpha_0)}{\prod_{i=1}^{k}\Gamma(\alpha_i)}\prod_{i=1}^{k}z_i^{\alpha_i-1}
\end{equation}
where $\alpha_0=\sum_{1}^{k}\alpha_i$. A symmetric Dirichlet distribution was also tested, meaning each parameter $\alpha_i$ has the same value. There is one more parameter, $\beta$, which has to be set properly. Such parameter assignment will be explained in the result section. Let $W_i$ be the $\text{i}^{th}$ topics, which is a vector of $N_i$ words long. The probability of a cluster $K$ generating a word $v$ at a position $j$ is

  \small
  \begin{gather} 
p(w_j= v | \alpha, \beta, L):\\
 \int_\Theta \sum_{k=1}^{K} \int_\theta p(w_j= v | \overrightarrow{\Theta}_k) p(z_j= k | \overrightarrow{\theta}_k) p(\overrightarrow{\Theta}_k | \beta)p(\overrightarrow{\theta} | \alpha) d\overrightarrow{\Theta}_k d\overrightarrow{\theta}_k
  \end{gather}
The goal is to maximize the probability $p(\theta, Z | W, \alpha, \beta, K)$.

\section{Result}\label{Result}
The implemented unsupervised learning method can help us discover hidden semantic structures. However, the number of topics should be determined and finding an optimal number of topics is challenging. To address this concern, perplexity and topic coherence were analyzed (Table \ref{ParameterComp}) \cite{PAVLINEK201783}. We considered our resultant topics to be sparse. Topic coherence was measured based on pointwise mutual information (PMI) by calculating word association between all pairs of words in all clusters.

\begin{equation}
PMI(W)=\frac{1}{\gamma}\sum_{i<j}^{}PMI(v_i,v_j), \text{     }i,j\in{W}
\end{equation}

\begin{equation}
PMI(v_i,v_j)=\log\frac{P(v_i,v_j)}{P(v_i)P(v_j)}
\end{equation}
where $\gamma$ is the number of PMI scores over the set of distinct word pairs.

Conditional log-likelihood of co-occurrence of top topic word pairs were also taken into account.

\begin{equation}
\zeta=\frac{2}{N(N-1)}\sum_{i=2}^{N}\sum_{j=1}^{i-1}\log\frac{P(v_i,v_j)}{P(v_j)}
\end{equation}
where $N$ is the number of top words in a cluster.

Hyper-parameters (i.e., $\alpha$ and $\beta$) affect the sparsity of topics and should be tuned properly. If a high value is assigned to the $\alpha$ parameter, various topics are generated. Hence, this value was set to a fraction of the number of topics. $\beta$ plays a role in the sparsity of words in the topics, and a high value means a lower impact of word sparsity. A low value means the topics should be more specific \cite{CHEN20171299,PANICHELLA2021106411}. Given the discussion, the $\beta$ parameter was set to $0.01$. We have selected the values such that maximum Coherence score and minimum perplexity for 5 clusters were obtained (Table \ref{ParameterComp}).

\begin{table}[ht]

\caption{Comparison of the model given different parameter setting} 
\centering 
\scriptsize
\begin{tabular}{l l c c c c c c c c c c c c} 
\hline\hline 

  \textbf{$\alpha$} &\textbf{ $\beta$}  & \textbf{Coherence-Value} & \textbf{Perplexity}
\\ [0.5ex]
\hline 
 
symmetric & 0.2 	& 0.260  & -7.626   \\[0.5ex]

\hline 

symmetric &	0.3		& 0.232  & -7.252    \\[0.5ex]
\hline 
0.1	& 0.2		& 0.257	 & -7.511
    \\[0.5ex]

\hline 
 
0.05	& 0.2		& 0.203 & -7.052
    \\[0.5ex]
\hline 
 
0.2	& 0.3& 0.195	 & -6.951
    \\[0.5ex]
\hline 
 
0.2 & 0.2	& 0.175 & -6.627
    \\[0.5ex]

\hline 

\end{tabular}
\label{ParameterComp}
\end{table}

\begin{figure*}
  \includegraphics[width=16cm]{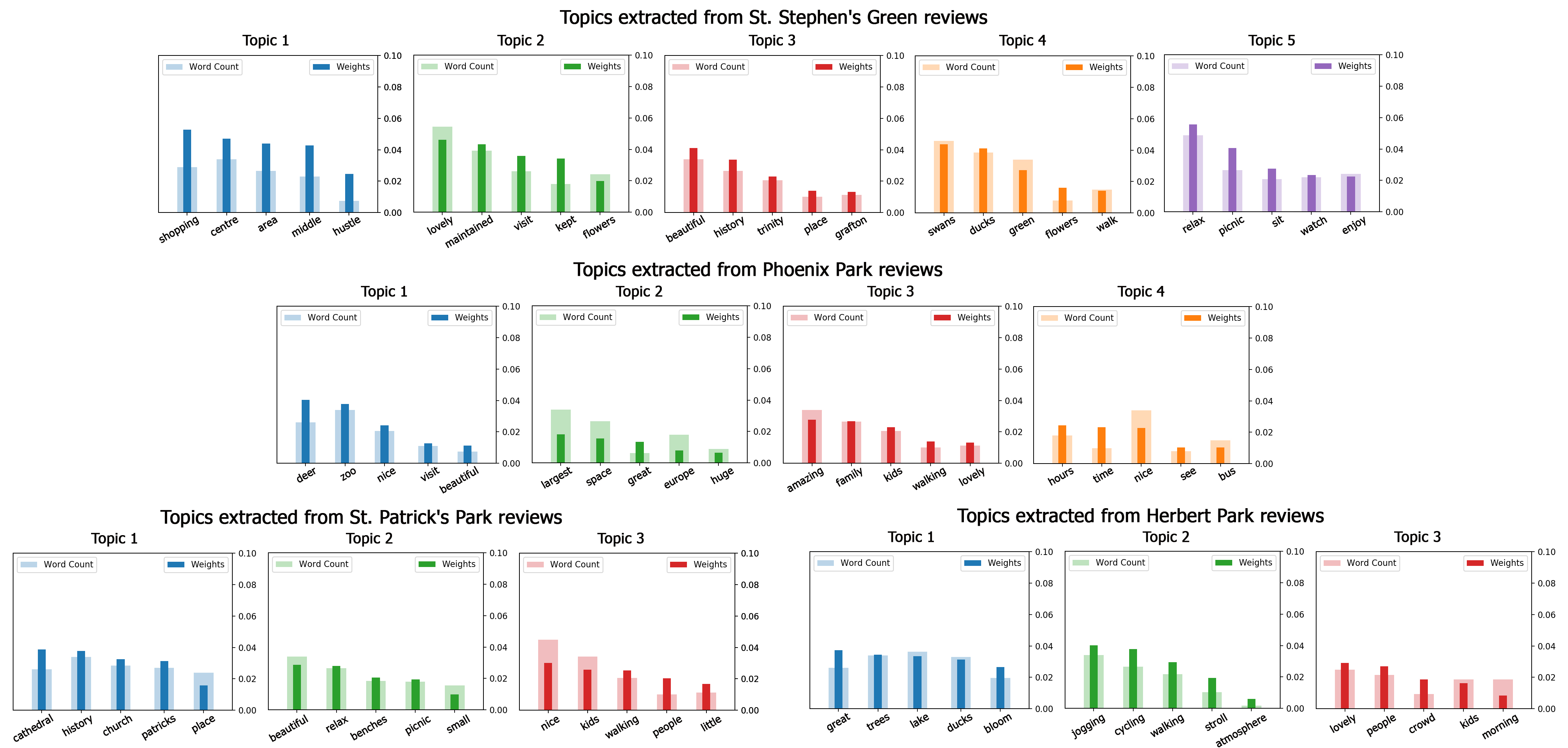}
  \caption{Extracted topics for 4 popular parks in Dublin, Ireland.}
  \label{topicsParks}
\end{figure*}

It should be mentioned that different datasets (related to most popular parks) were collected and the model was tuned separately. Extracted topics are a combination of keywords, each of which contributes a certain weightage to a given topic. Table \ref{keywordsTopics} reveals the importance (weights) of different keywords in the topics extracted from St. Stephen's Green reviews. Given the achieved keywords, Topic $1$ includes words like ''shopping'', "centre", "area", "middle", and "hustle", indicating a location-related topic. Topic $2$ includes words like "maintained" and "kept", which sounds like a management-related topic. Topic $3$ seems to be related to the historical background of the place. It includes words like "history" and "trinity" (Trinity College). Topic $4$ consists of words like "swans", "ducks", and "flowers", a wild and plant life-related topic. And Topic $5$ involves words such as "relax", "picnic", and "lunch", indicating a recreation-related topic. Fig. \ref{topicsParks} reveals the keywords associated to other most popular green spaces in Dublin. It should be mentioned that most popular parks have significant reviews, and the lesser common parks have fewer reviews. Therefore, the proposed method was used on the most popular parks as a proxy for UGS. The results indicate that user-generated content has the potential to drive insights about green space use and people's preferences towards green spaces. Such insights can be valuable for improving existing solutions and planning sustainable and socially equal cities.

\begin{table*}
\centering 
\scriptsize
\begin{tabular}{l c} 
\toprule 
\textbf{Topics } & Key characteristics of clusters and their corresponding weights\\ 
\midrule 
Topics \#1  & $\bullet$ "shopping": 0.053  $\bullet$ "centre": 0.044 $\bullet$ "area": 0.041 $\bullet$ "middle": 0.040  $\bullet$ "hustle": 0.023 $\bullet$ "grafton": 0.022 \\ %

Topics \#2   &  $\bullet$ "lovely": 0.046  $\bullet$ "maintained": 0.042 $\bullet$ "visit": 0.036 $\bullet$ "kept": 0.034  $\bullet$ "flowers": 0.020 $\bullet$ "walk": 0.018 \\ 

Topics \#3   & $\bullet$ "beautiful": 0.041 $\bullet$ "history": 0.035  $\bullet$ "trinity": 0.022 $\bullet$ "place": 0.014 $\bullet$ "grafton": 0.013  $\bullet$ "trees": 0.011 \\ 

Topics \#4   & $\bullet$ "swans": 0.042  $\bullet$ "ducks": 0.041 $\bullet$ "green": 0.027$\bullet$ "flowers": 0.016  $\bullet$ "walk": 0.014 \\ 

Topics \#5   & $\bullet$ "relax": 0.057 $\bullet$ "picnic": 0.041  $\bullet$ "sit": 0.029 $\bullet$ "watch": 0.023  $\bullet$ "enjoy": 0.022 $\bullet$ "lunch": 0.021 \\ 

\midrule 
\bottomrule 
\end{tabular}
\caption{Keywords for each extracted topic and associated weights given St. Stephen's Green reviews.} 
\label{keywordsTopics}
\end{table*}

\section{Conclusions and Future Work}\label{Conclusion}
Research to inform both policy and design of UGS is critical to protect these vulnerable areas while simultaneously ensuring access to the potential health and well-being benefits these spaces provide. Green spaces play a pivotal role across all aspects of city life, and as cities densify, the importance of accurately and effectively measuring the quality of UGS has never been greater.

Questionnaire surveys have been employed as a social science research tool to study people's preferences and their relationship with UGS. However, since data collection is an expensive and burdensome task, such studies are limited. To alleviate this concern, new digital data sources can be used. This work presents a novel approach to UGS management using two online platforms, i.e., Foursquare and TripAdvisor. The former was used to detect the most popular parks, and the latter was utilized to extract reviews associated with those parks identified. The relationship among the topics was also taken into account. Different validity tests were conducted to select optimal parameters. The proposed model can assess the quality of different urban environments cost-effectively, and hundreds of thousands of previously overlooked opinions can now be heard. The accessibility of the reviews makes the proposed method highly scalable—especially for popular parks. As discussed throughout the paper, Dublin follows a similar pattern as other cities, and most popular parks have significant reviews. Therefore, we suggest the proposed method to be used only on a city's most popular parks, as a proxy for UGS in cities, and then compare UGS between cities worldwide.

\bibliographystyle{elsarticle-harv}



{}


\end{document}